%% file: DocuByRefine.tex
\documentclass{llncs}
\usepackage{latexsym}
\usepackage{amssymb}
\ifx\pdfoutput\undefined
  \usepackage[dvips]{graphicx}
  \DeclareGraphicsExtensions{.eps}
  \usepackage[hypertex,bookmarks=false]{hyperref}
\else
  \usepackage[pdftex]{color}
  \usepackage[pdftex]{graphicx}
  \DeclareGraphicsExtensions{.pdf}
  \usepackage[pdftex,bookmarks=false]{hyperref}
  \hypersetup{colorlinks={true},
            linkcolor={blue},
            citecolor={blue},
            urlcolor={blue},
            plainpages={false}
  }
\fi

\input macros.tex

\begin{document}

\title{The Role of Executable Abstract Programs in Software Development and Documentation}
\author{Egon B{\"o}rger}
\institute{Universit\`{a} di Pisa, Dipartimento di Informatica,
           I-56125 Pisa, Italy
           \email{boerger@di.unipi.it;egon.boerger@gmail.com} 
           }

\maketitle

\begin{abstract}\footnote{This paper had been prepared for and accepted by the \emph{International Symposium On Leveraging Applications of Formal Methods, Verification and Validation} (ISoLA2022). Since it turned out that I could not present the contribution in person I decided to withdraw it from the conference Proceedings (a Springer LNCS volume).}
We present Executable Abstract Programs and analyse their role for software development and documentation. The intuitive understanding of these programs fits the computational mindset of software system engineers and is supported by a simple but precise behavioural definition. Therefore, they can be smoothly integrated in the practitioner's daily work to rigorously formulate every design and implementation decision taken on the path from the  Executable Abstract Program for the requirements to the targeted and efficiently runnable code. 

The Executable Abstract Programs of the resulting system documentation represent definitions of implementation steps one can check and justify by testing (due to their executable character) or by reasoning (due to the mathematical definition of their behaviour).
For complex systems the implementation involves multiple (orthogonal or successive) implementation steps which represent instances of a practical computational refinement concept. Such a system development process is driven by computational refinements and is strictly limited to explicitly formulate and justify---besides the requirements---only the necessary implementation steps. As a consequence, it produces as side-effect a corpus of documentation that facilitates the understandability of the final code and improves its reliability and resilience; it also enhances the maintenance process (including reuse and change of abstract programs and code) and reduces maintenance cost. 
\end{abstract}

\section{Introduction}\label{sect:intro}
Nobody would think that it could be viable, and for reasons of time pressure even desirable, to build a skyscraper without constructing a great variety of \emph{detailed documentation and mathematical model analysis} of a) what to build, b) what the environmental conditions that need to be respected are (static equilibrium, soil pressure, wind exposure, etc.), and c) how the building is realized: numerous blueprints of various parts (`components')---including their composition (`architecture')---and of different functions and views of the building, at different development stages, at different levels of detail, and with different purposes. And yet, still nowadays it is not rare to hear software engineers claim exactly that (just replace the word `skyscraper' by `software system')\footnote{This note 
	is not about simple algorithms or similar single-agent programs. The term `software system' refers to reactive software-intensive systems where multiple agents interact with each other and with a possibly physical environment in distributed concurrent (also called reactive) runs.}.
In Sect.~\ref{sect:methMisUndStand} we observe that historically this widespread attitude among practitioners may be the result of a too narrow understanding of `software specification' and its relation to code.\footnote{The read who is not interested in this purely theoretical and rather restricted understanding of the term `specification' may skip this section or come back to it after having see the notion of executable specifications we propose in Sect.3 to be used for practcial code development by stepwise refinements.} 

In Sect.~\ref{sect:specRefine} we explain the basic constituents of a practical notion of \emph{abstract yet executable programs} for software-intensive systems. The concept assigns a precise computational meaning to pseudo-code (hence the term `program') and can be used in practice to document via a checkable chain of implementation steps every design and coding decision taken during the code development. This documentation starts with a precise abstract yet executable program of the requirements (called \emph{ground model}, see Sect.~\ref{sect:whyExec}) and leads to compilable code. A ground model supports a rigorous inspection; that is, it makes it possible to check the \emph{appropriateness} of the executable abstract program with respect to the initial verbal or pictorial description of the requirements. All abstract executable programs support experimental validation (by testing) and mathematical verification (by proving desired system properties) of every single design decision. We point out that this improves not only the understandability and reliability of the code, but also enhances the maintenance process, including reuse and change of the abstract programs and the code, and reduces its cost. 

It is important that these abstract executable programs, in addition to the typical application domain notions that appear in the requirements, are formulated using only constructs and concepts that are familiar to the software engineer and in one way or another appear during any software development process: during the discussions in the minds of the designers and programmers, in their words or their sketches on the whiteboard, or as comments in the code. Due to their abstract nature, these constructs can be instantiated (we say `refined by implementation details') to express and document every design and coding decision taken during the development process. We explain why executable abstract programs are not specifications to be thrown away once the final code is available, but constitute an indispensable part of the documentation of the final code with which they must remain synchronized to efficiently support its understanding, the analysis of its well-functioning behaviour, and its maintenance. 
An authoritative quote affirms the practical relevance of such a code documentation:
\begin{quote}
\small
	A natural, comprehensive, and understandable description of the behavioural aspects of a system is a must in all stages of the system’s development cycle, and, for that matter, after it is completed too. \cite[p. 480]{HarPnu85}
\end{quote}

\section{Executable Abstract Programs are not Declarative}\label{sect:methMisUndStand}

The idea that software can be developed and maintained without an extensive---let alone rigorous---documentation (except the initial requirements and inline comments of the running code) seems to be  related to the widespread belief in the Formal Methods community that software specifications must be declarative, i.e. expressed by axioms, functional equations or logical formulae stating `what' the code should do but not `how'. We explain in this section why such specifications have only a limited range of applicability, often require to start again from scratch when it comes to building the desired runnable system, and thus not surprisingly are treated by many software engineers as a throw-away product: once the code is there the specification (if there is any at all) is not used anymore. Often it is even claimed that the code represents the only authoritative definition of the system so that specifications are not needed at all.

The bias toward declarative specifications as the good ones, as opposed to operational (the `distasteful') ones, is held by eminent and influential theoreticians and for decades has had and still has numerous followers in the Formal Methods community. To give just one outstanding example standing for many others, we quote what Tony Hoare stated recently in his contribution \emph{Forty years with Edsger} in \cite{AptHoa21}:
\begin{quote}
\small
	I shared his distaste for operational semantics. An axiomatic semantics can explain (maybe indirectly) the purpose of a concept and how it can be properly used (e.g., a chair is for sitting in). An operational semantics would have to give instructions for building a chair.
\end{quote}

The reader can find in \cite{Boerger94a} a precise technical criticism of this conception that underlies Dijkstra's understanding and use of formal methods for code design and verification. That criticism is still valid today. Hoare's example illustrates a characteristic point of this predilection for axiomatic statements, expressed in some fixed logic: the axioms specify properties of a static object (a chair) and its `proper use'. To see this in practical terms, replace `chair' with `software system' and consider as example Z \cite{Spivey92,DavWoo96}, one of the earliest and best-known formal methods. It is declarative, and is based on classical (first-order predicate) logic and the axiomatic set theory of Zermelo-Fraenkel. Z specifications typically describe programs statically, as abstract data types not involving any notion of run. This is considered to be `the most important characteristic of Z':

\begin{quote}
\small
	The most important characteristic of Z ... is that it is completely independent of any idea of computation. \cite[p.\ 89]{Hall97}	
\end{quote}

In fact, classical logic was devised (by Frege and others, long before the appearance of computing machines) to derive from axioms statements that are true in all Tarski structures (which represent the states of Z specifications). Action systems (in particular computer programs) have been devised to dynamically trigger changes of system states. The question is in which way logical statements that describe static states-of-affairs can be used  to describe dynamic state changes. Classical first-order logic is a logic for infinity, well-suited for standard mathematical investigations which are about static functions, predicates and relations in a world of infinite sets. But the world where programs run is finite and dynamic, in continuous evolution. Furthermore, to statically express a state change, declarative methods need some logical detour.

A simple example is the $x/x'$ notation of Z to declaratively describe that the value of $x$ is changed to a new value indicated by $x'$; the semantics of Z must guarantee that this value is then retrievable in the next state as the value of $x$. This triggers the frame problem of axiomatic descriptions, namely that one has to axiomatize in detail not only the relation between $x$ and $x'$ but also which variables $y$ (or more generally which memory locations) do NOT change their value. Since the new value is typically the value of some term $op(t)$, computed by applying a function or operation $op$ to the value $v$ of the term $t$ in the current state, a direct operational description of the state change that avoids the $x/x'$ notation and the frame problem uses a simple assignment $x:=op(v)$. If such abstract assignments $f(t_1,\ldots,t_n):=t$ (with arbitrary terms $t,t_1,\ldots,t_n$) because of their operational form are considered as `distasteful' and therefore should be avoided then, indeed, at least for the software developer, ``Z is in trouble" \cite[p.\ 89]{Hall97}.\footnote{Note that we are discussing the appropriateness of description methods for code development and not arguing against proof-theoretical investigations of program behaviour, using various logics, among them extensions of first-order logic for finite structures or modal logics for (dynamic logic) reasoning about computer programs. For a logic tailored to reason about Executable Abstract Programs see \cite{StaNan01,NanSta05,FeScTW17,FeScTW18}.} 

The idea that a specification should not be executable is held also outside the Z community; see the influential paper \cite{HayJon89} for arguments brought forward against the idea to execute specifications. We will come back to this in the next section and here remark only that the real issue is not any dichotomy between declarative (static) and operational (dynamic) descriptions, but the appropriate level of abstraction at which a description is given. If an operational description $M$ contains an abstract (e.g. declaratively defined) condition $cond$ or functional term $t$, for executions of $M$ for testing purposes one can replace $cond$ (resp. $t$) randomly by possible values (for $cond$ one of the two $true$ or $false$) without computing $cond$ (resp. $t$) in the simulated state. There is nothing wrong in providing an abstract declarative (typically a classical equational) definition for the input/output relation of a static mathematical function (for arguments and values of an infinite domain). 
Transformational methods as developed in functional or logic programming can be very helpful in obtaining code to compute such an equation or a logical axiomatization: if the specifying logical formulae are `close' to clauses of a target logic program or if the specifying equations are `close' to functions in the target functional programming language, one is not ``completely independent of any idea of computation"  \cite[p. 89]{Hall97}.  
But the situation changes when it comes to describing dynamical behaviour in a finite computational world. The following observations on Z apply to many transformational result-directed formal methods:

\begin{quote}
\small
	A Z schema can specify a functional system---one that produces an output in response to an input ... Z by itself is inadequate for specifying reactive systems. \cite{Lamport94} 
\end{quote}
\begin{quote}
\small
A reactive system, in general, does not compute or perform a function, but is supposed to maintain a certain relationship ... with its environment. ... Such systems do not lend themselves naturally to descriptions in terms of functions and
transformations. \cite[p. 479]{HarPnu85}
\end{quote}

Reactive systems exhibit patterns of collaborative actions of multiple agents working concurrently in (and stimulated by) a possibly complex computing and/or physical environment. The focus here is on how a state \emph{changes} by computational steps, and much less on invariants (properties that do not change during a run). In such a context the operational form allows one to \emph{directly} describe what does change due to an action, at its level of abstraction, the untouched parts remaining unchanged. Computational models for reactive systems lead directly to our theme in the next section, namely a computational concept of Abstract and Executable specifications that allows the practitioner to document every design and coding decision by a piece of (pseudo-) code.

\section{Documenting the Implementation of Abstractions by Executable Abstract Programs} \label{sect:specRefine}

In this section we are looking for a concept of Executable Abstract Program that can replace the widespread narrow interpretation of the term `specification'. The abstract character of such programs should allow the software engineer to refine them to directly reflect and thereby document any design or implementation idea---with the practical consequence of facilitating the understanding of the final code and of enhancing its maintenance process (including code changes and extensions) in terms of reliability and cost. 

\subsection{ASMs are Executable Abstract Programs}\label{sect:AEpgm}

The restricted applicability of purely declarative and not executable specifications discussed in Sect.~\ref{sect:methMisUndStand} leads us to define Executable Abstract Programs as \emph{abstract} and \emph{executable} pseudo-code. To complete this definition two things must be defined: 
\begin{itemize}
	\item the abstract objects, properties and relations---the abstract states---Executa- ble Abstract Programs operate on, 
	
	\item the abstract operations (their form and their computational meaning) that can be executed on those objects and form the pseudo-code. 
\end{itemize}

{\bf Abstract states}. Z made an appropriate proposal for the definition of abstract states: an abstract \emph{state} is given by a set $U$ together with a finite number of 
functions and relations defined on elements of $U$.\footnote{For notational simplicity but without loss of generality we represent relations $P$ by their characteristic predicate $C_P$ so that an atomic formula $P(x_1,\ldots,x_n)$ is viewed as an equation $C_P(x_1,\ldots,x_n)=true$ and states are given by a set $U$ together with finitely many functions defined on elements of $U$.} In mathematics they appear as models of theories, where the interest is to find statements that are true in a model. In logic they appear as Tarski structures and serve to semantically interpret formulae of first-order predicate logic (formalizations of statements). The models of mathematical theories and Tarski structures are static, and so are the abstract data types in traditional Z specifications. What is needed for computing is a \emph{dynamic} view of abstract states, i.e. of structures that evolve due to the application of operations---namely the pseudo-code operations defined below---to some of their objects, functions and relations.

{\bf Pseudo-code components}. An appropriate definition of abstract operations pseudo-code executes on elements of any state should directly reflect the dynamic change of the value of the involved objects, described by relational or functional terms, say change of the value of a function $f$ for arguments $t_1,\ldots,t_n$ to the value of a term $t$. This is what the fundamental computational 
\begin{center}
{\bf Assignment Operation: } $f(t_1,\ldots,t_n) := t$
\end{center}
does, abstractly for any terms and functions in any given state. This includes as special case the usual assignment of a value $v$ to an individual variable $x$ (viewed as 0-ary dynamic function) in programming languages, denoted $x:=v$. 

The assignment operation forms the basis of the following inductive definition of six further kinds of abstract pseudo-code operations. Software practitioners know these operations: they use them in their daily work with a well-known intuitive computational meaning that has a simple recursive definition (see \cite{BoeSta03,BoeRas18}) we do not repeat here.

\medskip
\noindent {\bf Bounded Parallel and Guarded Operations}:
\begin{itemize}
	
	\item a bounded parallel operation  \[\PAR (Op_1 ,\ldots, Op_n)\] executing \emph{simultaneously} finitely many already inductively defined operations $Op_1 ,\ldots, Op_n$; the term `bounded' refers to the fact that the number of simultaneously to be executed operations $Op_i$ is bounded by the parallel operation.\footnote{The $\PAR$ construct breaks with the prevailing sequential tradition of executing instructions one after the other. It turned out that the $\PAR$ construct helps to avoid any semantically irrelevant execution order, as for example in the $\ASM{Swap}(a,b)$ operation $\PAR(a:=b,b:=a)$. Offering a parallel operator to describe simultaneous instead of sequential execution helps to make independencies explicit that may become useful for parallel implementations.}
	
	\item a guarded (also called conditional) operation \[\IF \alpha \THEN Op\] where $\alpha$ is a Boolean combination of equations between first-order logic terms and $Op$ is an already inductively defined operation.
			
\end{itemize}
	    
\noindent 	{\bf Call Operations}\footnote{These two operations support the composition of complex operations out of simpler ones. They are definable in terms of sequential operations and substitution but for pragmatic reasons are included as basic operations.}:
	\begin{itemize}
	\item a call-by-value operation \[\LET x=t \IN Op\] where $Op$ is an already inductively defined operation where $x$ may occur and $t$ is a term,\footnote{This construct is also written $\WITH x=t \DO Op$.}
	\item a call-by-reference operation (also known as `call-by-name') \[Op(t_1,\ldots,t_n)\] to trigger the execution of an instance (with terms $t_1,\ldots,t_n$) of an operationdefined by a declaration of the form $Op(x_1,\ldots,x_n)=Q$, where $Q$ is an already inductively defined operation. 
\end{itemize}
\noindent  {\bf Logic Operations $\CHOOSE / \FORALL$}:
\begin{itemize}
	\item an unbounded\footnote{The term `unbounded' refers to the fact that the number of elements $x$ involved in the $Op$eration is not bounded by the size of the program but by the possibly infinite set of elements satisfying $\alpha$ for which the $Op$eration is executed.} parallel operation \[\FORALL x \WITH \alpha \DO Op,\]  where $\alpha$ is an arbitrary (usually first-order) condition and $Op$ is an already inductively defined operation where $x$ may occur,
	
	\item an unbounded choice operation \[\CHOOSE x \WITH \alpha \DO Op\] to describe non-deterministic behaviour,\footnote{The $\CHOOSE$ operator is typically used to abstract from particular scheduling mechanisms when selecting an element to execute an operation. Similarly to the $\PAR$ operation,  $\CHOOSE$ helps to make it explicit (and thereby separately refinable by different selection strategies) when any execution order for the possible choices yields the desired behavioural effect. }  where $\alpha$ is an arbitrary  (usually first-order) condition and $Op$ is an already inductively defined operation where $x$ may occur.
\end{itemize}

If one works in a special application domain, it is often an advantage to add to these seven constructs some fundamental domain-specific operations with appropriately described behavioural meaning, thereby turning Executable Abstract Programs into a customized Domain-Specific Language (DSL). 
\medskip

\noindent {\bf Abstract State Machines}. Each of the above seven general pseudo-code operations describes an Abstract State Machine (ASM \cite{BoeRas18,BoeSta03}). Therefore, we identify Executable Abstract Programs with ASMs. For the formal definition of the intuitive semantics of ASMs explained above, we refer to the textbooks and mention here only that multi-agent ASMs---any finite set of agents each equipped with a pseudo-code operation as its program (called also ASM rule)---are reactive systems where each single-agent component ASM executes a loop of its rule.

Since the late 1980s, the operations listed above have been used with success to faithfully model all kinds of computational systems,\footnote{See \cite{Boerger02a} for an early survey (until 2003) and for recent work \url{https://abz-conf.org/method/asm/} and \cite{BoeRas18}.} which highlighted the practical industrial-strength usefulness of these Abstract Executable Programs.

{\bf Epistemological side remark}. The operations also have an interesting epistemological property, more specifically concerning the concept of algorithms. Let {\bf PGA} (standing for {\bf Parallel Guarded Assignment} machine) denote ASMs one can define applying only assignment, bounded parallel and guarded operations. PGAs characterize `sequential algorithms' in the sense of the axiomatic description of the concept in \cite{Gurevich00}.\footnote{As a result, in the literature PGAs are called sequential ASMs although their definition invites to think in terms of parallel execution wherever behaviourally possible.} The characterization implies that PGA operations have a simple normal form, namely $\PAR (Op_1,\ldots,Op_n)$, where each $Op_i$ is a conditional operation of form $\IF cond_i \THEN assign_i$ with guard $cond_i$ and an assignment operation $assign_i$. 
Other classes $\mathcal{S}$ of computational systems---synchronous parallel \cite{FeScTW16}, concurrent (reactive)
\cite{BoeSch16}, recursive \cite{BoeSch20}, reflective \cite{SchFer20}, etc.---could be characterized in terms of corresponding classes  $ASM(\mathcal{S})$ of ASMs, defined by constraints on their states and rules. More precisely, an axiomatic machine-independent definition $Def(\mathcal{S})$ of elements of $\mathcal{S}$ is provided together with a definition of a class $ASM(\mathcal{S})$ of ASMs and a proof that a) every machine $M$ in  $ASM(\mathcal{S})$ satisfies the axioms $Def(\mathcal{S})$, and b) for every system that satisfies the axioms $Def(\mathcal{S})$ its runs can be simulated step-by-step by the runs of a corresponding ASM in $ASM(\mathcal{S})$. 

\medskip 
Note that the combination of abstractness with executability is an intellectual concept of performing runs of state changes by applying in each step abstract operations to abstract objects; this run concept fits a computational mindset and becomes a purely mechanical feature by an appropriate implementation (typically involving some refinement  steps, see Sect.~\ref{sect:AErefine} below) that translates abstract operations into concrete physical machine operations.

Before explaining how such a \emph{practical refinement relation} supports an efficient code documentation, let us consider some main reasons for the desirability of the executability of abstract models.

\subsection{Reasons for Executability}\label{sect:whyExec}

There are many reasons to require the executability of abstract programs. A pragmatic one is that a program that runs fits the computational mind of the engineers; not surprisingly, today a large community of practitioners use executable UML \cite{MelBal02} for software specifications.\footnote{ASMs add rigour to UML notations, see \cite{Sarstedt06,SarGut06,KohGut09,Kohlmeyer09}.} We mention here only three more reasons that are related to the documentation issue. 

A practical reason is due to the fact that the requirements may be given only in terms of scenarios, describing example behaviours the code is expected to handle. As an example we can mention the industrial FALKO project \cite{BoPaSc00}, where the requirements for a railway timetable validation and construction program were given by a set of concrete schedules and data of train runs in two major German local transportation areas; the goal was to design and implement the railway process model component in such a way that it could handle the scenarios indicated. Various ASM models were built and extensively tested using the ASM Workbench
\cite{DelCastillo00} to run the scenarios before compiling the final model using an Asm2C++ compiler \cite{Schmid01}. The final ASM model was documented as a precise model of the requirements and as an abstract model of the C++ code. This allowed us to reuse the ASM model when a change request appeared: we refined the model to integrate the additional requirement and recompiled the new model so that the new code was again in sync with the refined requirements model. In such maintenance and change usage of models (not of the final code) it is important, as one 
reviewer observed, that the generated code is ok as generated, without need of any further ‘improvements’ (`optimizations').
The program worked without failure for years in the Vienna local transportation system, and no maintenance cost was incurred.

A second reason for executability of abstract programs is to use them for rapid prototyping, i.e. as prototypical or as reference implementation of a system. As recent example we can point to the well-known Subject-oriented Business Process Modeling (S-BPM) approach where the ASM interpreter CoreASM \cite{CoreASM} is used for a reference implementation \cite{WoBoHe19,WoBoHe19a} of an industrial-strength  workflow engine programmed in $C^\#$ (see \cite[p. 298]{ElWFSB22}).  

A third, and perhaps the most important, methodological reason for executability of abstract programs is that it provides a basis for early correctness checks by testing (validation), long before reaching the detailed level of compilable code. It is well-known that abstraction helps to find conceptual problems that are more difficult to detect in compiled code and typically remain unrecognized by standard debugging techniques. Executability helps to document at a high level of abstraction the conceptual role of dynamic concepts. This holds in full generality and is well illustrated by an example from a famous author's hand, where a well-known Termination Detection Protocol was developed ``to demonstrate how the algorithm can be derived in a number of steps" \cite{Dijkstra83}. When this algorithm was chosen in a Dagstuhl seminar\footnote{Seminar 13372 on ``Integration of Tools for Rigorous Software Construction and Analysis", 2013.} for a formalization test with various formal methods, it turned out that exactly the operational character of the ASM model \cite{GerRic14}, which permitted to run it in CoreASM \cite{CoreASM} (or in ASMeta \cite{Asmeta}), revealed a certain number of problems concerning the correctness of the algorithm; typically these were hidden assumptions that also other declarative specifications (produced during the seminar) made without noticing.\footnote{For a detailed exposition see \cite[Ch.3.2]{BoeRas18}.} 

\subsubsection{Validation of Ground Models}\label{sect:groundValid}
An important subcase of model validation concerns what we call \emph{ground models} \cite{Boerger03a}. These models constitute a precise and complete description of the requirements (and of the domain assumptions the software developers can rely upon), i.e. of the system behaviour as seen at the level of abstraction of the application domain. Thus, a ground model explains how the intended system behaviour ``relates to the affairs of the world it helps to handle" \cite[p. 254]{Naur85}. By this (epistemological, not formal!) definition, ground models cannot be proved (in the mathematical sense) to be correct and complete, given that they represent the only initial  precise description of the requirements. But \emph{inspection} can help provide confidence in the model's correctness and completeness---if the model is formulated in terms both parties understand: the application domain experts and the software designers.\footnote{The understandability requirement is satisfied for ASMs by their form---the constructs are close to natural-language expressions---and by their natural intuitive meaning that has a simple  precise foundation, see the seven constructs in Sect.~\ref{sect:AEpgm}.} 

This comprehensibility of ground models implies that the language in which they are formulated must embrace a most general notion of state and state change (`abstraction' concept) so that the inspection can rely upon a direct (in Aristotelian terms an \emph{evident}) correspondence between the application domain real-world-view and the ground model (i.e. the application domain) concepts. A reviewer pointed out a to-be-mentioned instance where such a correspondence appears: when domain features or concepts are modeled using (hierarchies of) types or classes, these types and classes do appear in the ground model. A characteristic example are the various submodels of the ground model ASM $\ASM{Java}$ defined in \cite{BoeSta03}; notably, most of their description is about static features concerning types, classes and their structural (hierarchical) relations, the crucial dynamic features constitute only a small part of the entire abstract executable ASM model. In the concurrent ASM $\ASM{JavaThread}$ (where every agent executes an instance of the ASM $\ASM{Java}$) the  universe $Thread$ represents the domain of objects that belong to the Java class Thread. 
In a recent doctoral dissertation \cite[Ch.5]{Paulweber22} a new ASM construct is described and implemented that allows one to define a new type abstraction by specifying a signature-and-rule pair; this permits to modularize ASMs.

Furthermore, for validation purposes the inspection process must be supported by the possibility to run the ground model, whether intellectually or mechanically (by some execution mechanism). ASMs satisfy this condition. The behaviour-driven development (BDD) method \cite{BDDmethod} even advocates the use of executable specifications to provide a common ground for application domain experts and software designers to agree on what to build.\footnote{As observed in \cite[Sect. 4.4]{ABBGRS21}, the ASMeta toolset offers to generate C++ executable BDD scenarios from ASM (scenario) models.} To play its role well, the ground model must be fully documented because it must be in sync with the final code, a property that must be maintained in every implementation step.\footnote{In successful systems, requirements and the implementing code may evolve, but each time one set of requirements is refined to one implementation, which for brevity we call `final code'.}  This leads us to explain the role of refinement for documentation.

\subsection{ASM Refinements to Implement Abstractions}\label{sect:AErefine}

{\bf Scientific purpose of ASM refinements}. The idea to develop programs by a series of refinement steps is over 50 years old by now (see \cite{Wirth71,Dijkstra72}). The definition of ASM refinements in \cite{Boerger02b} generalizes and simplifies other (mostly syntactically or proof-theoretically) restricted refinement concepts. It exploits the operational and abstract character of ASMs which allows the practitioner to formulate at any level of abstraction implementation steps that introduce details for some abstract data and, where necessary, also for some abstract operations. Among others it includes vertical refinements (implementations of some abstract data or actions) and horizontal refinements (extensions by introduction of new functionality, typically involving new data and new actions).
The definition of ASM refinements is made in such a way that \emph{any implementation idea can be} {\bf directly} \emph{reflected by an ASM refinement} step. This close relation between an abstract program $M$ and a more detailed program $M^*$ allows the practitioner to justify the implementation step that led from  $M$ to  $M^*$ by checking that corresponding runs of $M$ and  $M^*$ are in the desired behavioural implementation correctness relation.\footnote{See diagram 2.1 (p. 24) in \cite{BoeSta03} and the definition on p. 111 we do not repeat here.} 

Declarative models offer a special refinement use. For example, a functional definition $f(x_1,\ldots,x_n) = t$ allows one to separate the interface use of $f$ from its implementation. However, in such a definition the data are fixed. What ASM refinements add is that also the data, not only the operations, can be refined together. Program libraries for static or dynamic functions support executable ASMs where these functions are part of the background.

{\bf Documentation role of ASM refinements}. When using ASMs, the justification that an implementation step is `correct'---i.e. that runs of the refined machine $M^*$ reflect the corresponding runs of the more abstract machine $M$ in the intended way---can be given by an experimental validation (exploiting the executability) or by a mathematical verification (exploiting the mathematical character of ASMs), or both. The important point is that $M$ and $M^*$ are not throw-away products but part of the documentation of how the specific implementation of (some of the concepts and behavioural features of) $M$ by $M^*$ contributes to the overall implementation, namely of the requirements (as specified in the ground model) by the to-be-developed compilable code. The documentation of the---in complex cases anyway stepwise\footnote{No work is done twice, every refinement step describes a concrete implementation step}---implementation process supports the important software engineering principles known as design-for-change and design-for-reuse: changes can be easily obtained at any level of abstraction the implementation went through, just change $M^*$! 
Obviously, the documentation of (the set or sequence of) implementation steps also facilitates the comprehension (often called `readability') and the analysis of the final code (if the intermediate programs are kept in sync with it). It also reduces the cost of software maintenance. These are two pragmatic reasons that economically justify the effort---really an intellectual discipline---to keep the documentation and in particular the relation between the ground model and the final code consistent. When evaluating the development cost, the cost for the documentation should be compared not only with the cost for maintenance, but also with the cost system failures may trigger: cost of downtime and recovery work, cost of safety violations, cost of damage by security breaches, etc.

{\bf Scalability}. The notion of ASM refinement as expression of implementation steps proved capable of scaling to industrial-size systems, supporting the splitting of a complex design task into the design of an (often hierarchically structured) set of simpler, piecemeal testable or verifiable and reusable components. Refinement is thus a pillar of the ASM method for rigorous system design and analysis \cite{BoeSta03,BoeRas18} and via the documentation of the ground and the intermediate models supports design-for-change, design-for-reuse, and component-based system engineering. To trigger the reader's interest we mention two industrial examples of hierarchies of implementation steps, handled (specified, tested and verified) as ASM refinements. 

{\bf The first example} illustrates the use of ASMs to test and verify programming language and compiler properties. In \cite{BoeDas90} (streamlined in \cite{BoeRos95a}), an ASM tree model for the ISO standard of Prolog was defined by a core ASM for user-defined predicates together with successive stepwise refinements by components for control constructs, database operations, solution collecting predicates, error handling features, and the box model for debugging.\footnote{In \cite{Kappel90,Kappel93} the world-wide first machine to physically execute sequential ASMs was implemented in Prolog and extended in \cite{BCFHKL96} by a stream-based communication parallelism  for a time- and security-critical coal mining application in Germany. The same year D. Bowen \cite{Bowen91} implemented B\"orger's Prolog model in \cite{Boerger90a} and used it at Quintus for experiments.} In \cite{BoeRos95} this machine was further refined to a stack model (with choicepoint reuse), to which successively the following implementation steps were applied to reach the level of WAM code\footnote{\url{https://en.wikipedia.org/wiki/Warren_Abstract_Machine}}: predicate structure compilation (look-ahead optimization, switching), clause structure compilation (sharing the continuation stack), term structure compilation (heap, push-down list, unification, putting, getting, substitution and trail stack for bindings, environment trimming, last call optimization), and WAM (temporary variable optimizations and cut). For every refinement step a precise formulation of what its correctness means was given and was proven in \cite{BoeRos95} to hold. Later this proof was mechanically verified using the KIV theorem prover \cite{SchAhr97,SchAhr98}. However, such a mechanical verification comes at a cost. A software engineer may well be able to `justify' his design steps by appropriate (possibly mathematical) arguments, but turning them into machine-supported proofs in some logic may need a deep expertise with the underlying (implementation of) logic and therefore require help from a logician or mechanical theorem proving specialist.  For example, the work in \cite{BoeRos95} to refine the given Prolog specification to WAM code and to mathematically prove the correctness of the refinement steps took 6 person-months, whereas the refinement to KIV took 24 person-months of work. This example shows that, from the engineering point of view, for such a complex enterprise it is better to separate the two steps involved: software engineers design and verify the system (here the models for Prolog, the WAM and the compiler), while verification experts enrich the specification and proof steps with those details the prover needs to know to handle the proofs. 

{\bf The second example} is about the reuse of ASMs. The Prolog and WAM ASMs in \cite{BoeRos95a,BoeRos95} were reused to define interpreters for various other logic programming languages, including IBM's languages PROTOS-L \cite{BeiBoe96a,BeiBoe96b} and $CLP(\cal R)$ \cite{BoeSal94} (see \cite[p. 346-347]{BoeSta03} for more references).
In \cite{StScBo01} operational models for Java and for the JVM were developed by stepwise refinement\footnote{The definition works even instructionwise and can be used in connection with software product lines, as has been observed in \cite{BatBoe08}.} together with a stepwise defined and proven-to-be-correct compilation scheme of Java programs to JVM bytecode. These interpreters come with executable AsmGofer \cite{Schmid99,Schmid02} versions that have been extensively used for testing.\footnote{AsmGofer has been used also in an industrial ASIC design and verification project at Siemens, together with a compiler from ASM to VHDL.} By comparing test runs of the ASMs (for Java, the compiler and the JVM) with runs of Sun's machines, we discovered various bugs, all of which were reported to and corrected by Sun.\footnote{Practitioners may be interested to see some notable details to be found in the appendix.} The Java model was reused, on the request of Microsoft Research Cambridge, to define the ECMA semantics of $C^{\#}$ (see \cite{BoFrGS03}) and later to define an ASM model for the semantics of an object-oriented programming language of which both Java and $C^{\#}$ are refinement instances, see \cite{BoeSta04}.

For numerous other examples of hierarchies of stepwise-refined executable ASMs see \cite{BoeSta03,BoeRas18} and \url{https://abz-conf.org/method/asm/}.

\medskip

{\bf Tool support}. ASM refinement chains can be documented using some standard Version Control (VC) system in combination with an appropriate ASM editor and a cooperative work supporting tool. VC systems provide sequences of backtrackable project states. A project-history-focussed version management can be enhanced by taking precise behavioural criteria into account: define the project's ground model and each ASM refinement as a separate version. Then possible fork (and correspondingly merge) points document different implementation steps of some intermediate machines, at some intermediate level of abstraction. This allows the developers to move between high-level and low-level system views, in both directions along the refinement hierarchy, to examine details, to operate changes, to continue the development with further implementation steps, etc. Such a use of VC produces also a structural big-picture of the behavioural interdependence of the components of the system under construction.  

Also ``Agile methods" and ``Extreme Programming" (XP) can be used and enhanced, defining the incremental `intermediate' programs (read: still abstract but executable programs) by a practical but rigorously analysable refinement concept (`practical specifications of design decisions'), as offered by ASMs and their refinement notion. As a result the program increments are linked in a behaviourally consistent way one can verify (mathematically) and validate (experimentally by testing methods).

\medskip

{\bf Iterative refinement character}. It is a well-known experience that the development process that leads from the requirements (read: a ground model) to compilable code is by no means linear.

\begin{figure}
	\centering
	\includegraphics[width=0.9\linewidth]{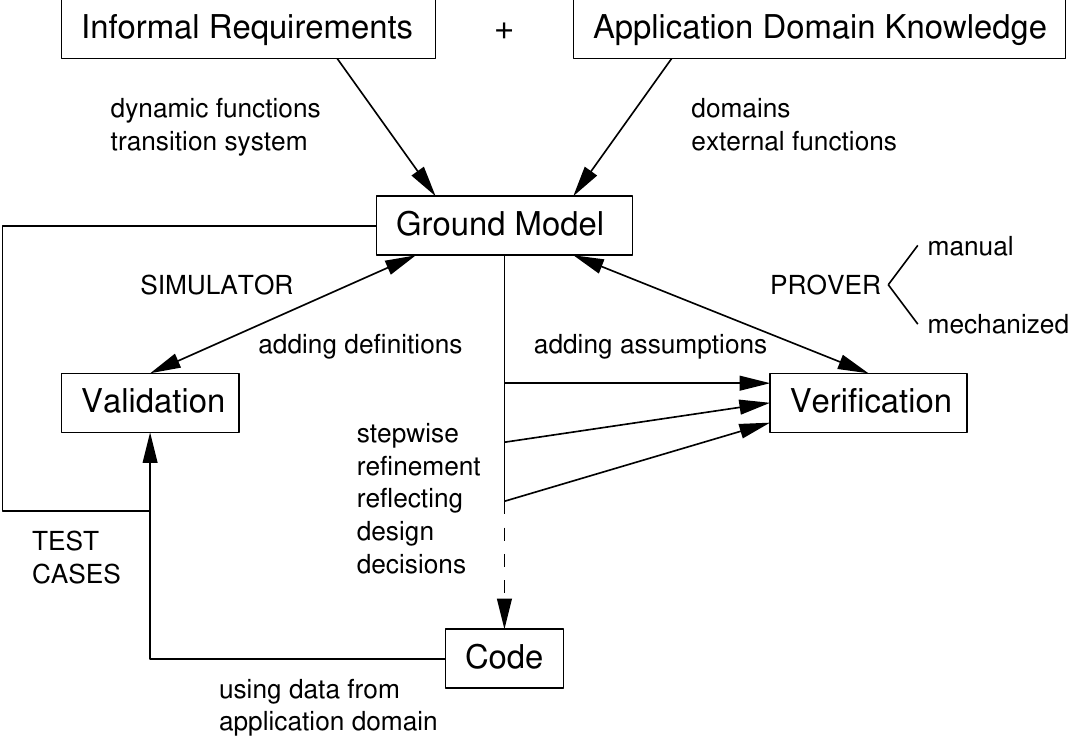}
	\caption{Iterative Character of ASM Development Method}
	\label{fig:overview}
\end{figure}

An implementation step may reveal that some condition is missing or not stated clearly enough in the requirements. Or the customer formulates a new requirement to be included, a phenomenon known as evolution of requirements (see the FALKO example mentioned in Sect.~\ref{sect:whyExec}). In such cases the customer must be involved to clarify the issue, leading to a modified ground model and corresponding changes in some refined models that are affected by the change; the documentation serves to precisely identify the locations involved and to keep track of the additions or changes, instead of erasing after the discussion what had been written on the whiteboard. The same can happen with respect to intermediate levels of implementation and makes the programs resilient. The back and forth is needed to keep the intermediate implementation models in sync with the final compilable code. This does not mean that a specification is written twice, it means that during the development process it turned out that the understanding of what to build was not complete and had to be made more precise. Thus, the documentation helps detect mistakes as early as possible. 

To conclude we illustrate by Fig.~\ref{fig:overview} (taken from \cite{BoeSta03}\footnote{\textcopyright{~2003 Springer-Verlag Berlin Heidelberg, reused with permission}.}) the iterative, not necessarily linear, character of developing a system by stepwise implementation. Stepwise refinements establish a kind of flight of stairs that enables walking between different levels of abstraction; a typical reuse application creates a new branch at the level of abstraction where the change is formulated. Only very skilled tightrope walkers can fly between multiple levels of abstraction without securing their balance via some concrete and well-documented  abstraction level where to put one's feet for the dance from requirements to binary code. 

\section{Conclusion}
We have explained why Abstract State Machines, a rigorous form of  \emph{Executable Abstract Programs}, can help the practitioner in his daily work to rigorously formulate and document one by one all design and implementation decisions that bridge the gap between requirements and efficiently runnable code. To evaluate the quality of the documentation produced by the stepwise design we refer the reader to the indicated 
projects. More of that can be found on the ASM website \url{https://abz-conf.org/method/asm/} and in the bibliography of the two books \cite{BoeSta03,BoeRas18}. We would be interested if somebody
knowledgeable about the unified specification/modeling and programming approach \cite{BrHaKu16} compares it with the ASM method.

\medskip

{\bf Acknowledgment}. Thanks to the following colleagues who generously helped with critical comments and suggestions: Don Batory, Paolo Dini, Flavio Ferrarotti, Albert Fleischmann, Uwe Gl\"asser, Philipp Paulweber, Alexander Raschke, Elvinia Riccobene, Klaus-Dieter Schewe and two anonymous referees who accepted the paper for presentation in and publication by the 
\emph{International Symposium On Leveraging Applications of Formal Methods, Verification and Validation} (ISoLA2022).

\section{Appendix: Mixing Runs of AsmGofer/Sun Machines}

\begin{figure}
	\centering
	\includegraphics[width=0.8\linewidth]{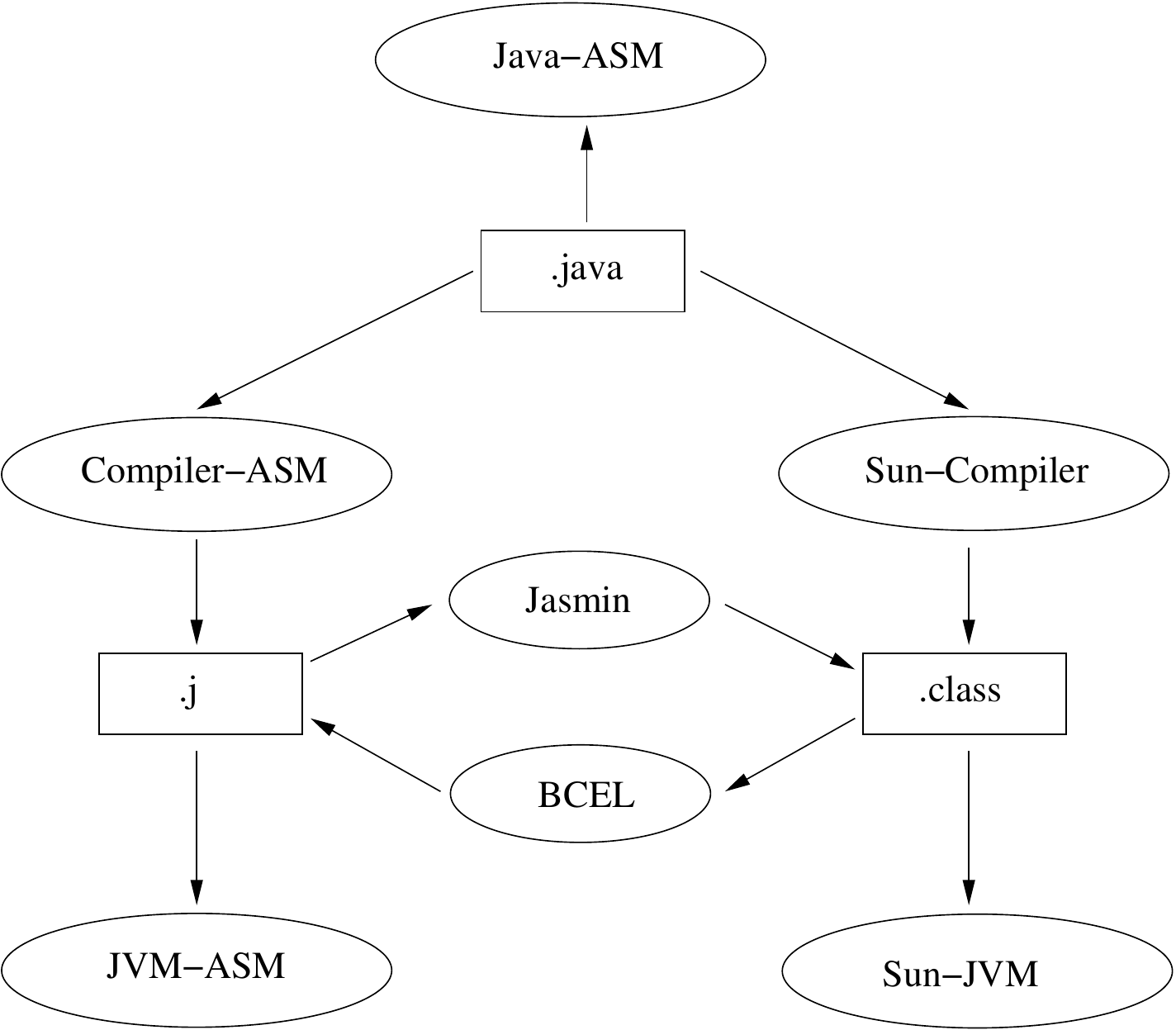}
	\caption{AsmGofer and Sun Machines for Java/JVM}
	\label{fig:JBookFiletypes.pdf}
\end{figure}

Figure~\ref{fig:JBookFiletypes.pdf}  (taken from \cite{StScBo01}\footnote{\textcopyright{~2001 Springer-Verlag Berlin Heidelberg, reused with permission}.}) shows how the AsmGofer machines for Java/JVM (i.e. mechanically executable refinements of the ground models for Java and the JVM developed in the Jbook \cite{StScBo01}) can be run together with Sun's JVM, using both our and Sun's compiler.

For better readability we adopted a textual representation of the ASMs, using converters Jasmin (from textual to binary) and BCEL (from binary to textual).

\begin{itemize}
	\item Java-ASM reads and executes Java source code (suffix .java). Sun has no counterpart for this ground model. We submitted it to a careful inspection, using the Java/JVM manuals as description of the  requirements, and added an extensive validation by examples whenever an unclear situation showed up.
	\item Compiler-ASM compiles from source code .java to bytecode .j in Jasmin syntax (textual representation).
	\item Sun-Compiler generates .class files (binary class file representation).
	\item JVM-ASM resp.Sun-JVM read input in Jasmin syntax resp. class files.
\end{itemize}
For the validation of the AsmGofer machines and our compiler, using the converters we could mix the different machines and compilers and compare the runs. Any inconsistency revealed by the comparison showed a bug either in our definitions or in Sun's code.

\def\note#1{}

\bibliographystyle{abbrv}

\bibliography{DocuByRefine}

\end{document}

%% file: macros.tex
\makeatletter
\def~{\ifmmode\;\else\penalty\@M\ \fi}
\def\@setmcodes#1#2#3{{\count0=#1 \count1=#3
  \loop \global\mathcode\count0=\count1 \ifnum \count0<#2
  \advance\count0 by1 \advance\count1 by1 \repeat}}
\DeclareSymbolFont{italic}{OT1}{\rmdefault}{m}{it}
\let\mathit\undefined
\DeclareSymbolFontAlphabet{\mathit}{italic}
\edef\@tempa{\hexnumber@\symitalic}
\@setmcodes{`A}{`Z}{"7\@tempa41}
\@setmcodes{`a}{`z}{"7\@tempa61}
\makeatother
\newdimen\asmindent     
\asmindent=\parindent
\newcount\asmi
\def\inc{\global\advance\asmi by 1}
\def\dec{\global\advance\asmi by-1}
\def\nl{{}$\par\hangindent\asmi em
  \noindent\kern\asmi em\ignorespaces$} 
\def\asmskip{{}$\par\smallskip\hangindent\asmi em
  \noindent\kern\asmi em\ignorespaces$} 
\def\asm{\global\asmi=0 
 \def\+{\inc\nl}
 \def\-{\dec\nl}
 \def\\{\nl}
 \begin{trivlist}\item[]\leftskip=\asmindent\relax$}
\def\endasm{$\end{trivlist}}
\def\asmarray{\begin{array}[t]{@{}l@{\;}l@{\;}l@{}}}
\def\endasmarray{\end{array}}
\newcount\asmii
\def\subasm{\vtop\bgroup\asmii=0\normalbaselines
 \def\nl##1{$\egroup\advance\asmii by##1\relax\hbox\bgroup\hskip\asmii em$}
 \def\\{\nl{0}}
 \def\+{\nl{1}}
 \def\-{\nl{-1}}
 \hbox\bgroup\hskip\asmii em$}
\def\endsubasm{$\egroup\egroup}
\def\ASM#1{\hbox{\sc#1}}        

\def\CHOOSE  {\mathrel{\mathbf{choose}}}

\def\DO      {\mathrel{\mathbf{do}}}

\def\FORALL  {\mathrel{\mathbf{forall}}}

\def\IF      {\mathrel{\mathbf{if}}}

\def\IN      {\mathrel{\mathbf{in}}}
\def\LET     {\mathrel{\mathbf{let}}}

\def\PAR     {\mathrel{\mathbf{par}}}

\def\THEN    {\mathrel{\mathbf{then}}}

\def\WITH    {\mathrel{\mathbf{with}}}

\makeatletter
\def\enumerate{%
  \ifnum \@enumdepth >\thr@@\@toodeep\else
    \advance\@enumdepth\@ne
    \edef\@enumctr{enum\romannumeral\the\@enumdepth}%
      \expandafter
      \list
        \csname label\@enumctr\endcsname
        {\usecounter\@enumctr\def\makelabel##1{\hss\llap{##1}}
         \itemsep 0pt\parskip 0pt\parsep 0pt\topsep\smallskipamount}%
  \fi}
\def\itemize{%
  \ifnum \@itemdepth >\thr@@\@toodeep\else
    \advance\@itemdepth\@ne
    \edef\@itemitem{labelitem\romannumeral\the\@itemdepth}%
    \expandafter
    \list
      \csname\@itemitem\endcsname
      {\def\makelabel##1{\hss\llap{##1}}
       \itemsep 0pt\parskip 0pt\parsep 0pt\topsep\smallskipamount}%
  \fi}
\makeatother